\documentclass[sn-mathphys,Numbered]{sn-jnl}% Math and Physical Sciences Reference Style

\usepackage{algorithm}%
\usepackage{algorithmicx}%
\usepackage{algpseudocode}%
\usepackage{amsmath,amssymb,amsfonts}%
\usepackage{amsthm}%
\usepackage{anyfontsize}
\usepackage[title]{appendix}%
\usepackage{booktabs}%
\usepackage{graphicx}%
\usepackage{listings}%
\usepackage{lmodern}
\usepackage{manyfoot}%
\usepackage{mathrsfs}%
\usepackage[version=4]{mhchem}
\usepackage{multirow}%
\usepackage{textcomp}%
\usepackage{ulem}
\usepackage{units}
\usepackage{xcolor}%
\begin{document}

\title[title]{Ultrathin 3R-\ce{MoS2} metasurfaces with atomically precise edges for efficient nonlinear nanophotonics}

\author*[1]{\fnm{George} \sur{Zograf}}\email{georgii.zograf@chalmers.se}

\author[1]{\fnm{Bet$\textrm{{\"u}}$l} \sur{K$\textrm{{\"u}}$$\textrm{{\c{c}}}$$\textrm{{\"u}}$k$\textrm{{\"o}}$z}}

\author[1]{\fnm{Alexander Yu.} \sur{Polyakov}}

\author[2]{\fnm{Maria} \sur{Bancerek}}

\author[1]{\fnm{Abhay V.} \sur{Agrawal}}

\author[3]{\fnm{Witlef} \sur{Wieczorek}}

\author[1,2]{\fnm{Tomasz J.} \sur{Antosiewicz}}

\author*[1]{\fnm{Timur O.} \sur{Shegai}}\email{timurs@chalmers.se}


\affil[1]{\orgdiv{Department of Physics}, \orgname{Chalmers University of Technology}, \postcode{412 96}, \city{G$\textrm{{\"o}}$teborg}, \country{Sweden}}

\affil[2]{\orgdiv{Faculty of Physics}, \orgname{University of Warsaw}, \orgaddress{\street{Pasteura 5}, \postcode{02-093}, \city{Warsaw}, \country{Poland}}}

\affil[3]{\orgdiv{Department of Microtechnology and Nanoscience}, \orgname{Chalmers University of Technology}, \postcode{412 96}, \city{G$\textrm{{\"o}}$teborg}, \country{Sweden}}


\abstract{%Nonlinear optical phenomena, such as light modulation, quantum photons generation, and extreme nonlinear enhancement of the laser intensities through optical parametric amplification, rely heavily on inherent material properties of the active media. Conventional methods of improving the nonlinear process efficiencies involve selecting materials with high nonlinear susceptibility ($|\chi^{(2)}|$) and cavities with high-quality factors to concentrate electromagnetic field in the active media volume. All-dielectric nanophotonics provide a platform for self-consistent nonlinear optical systems, where the resonant structure serves both as an active nonlinear media and a compact cavity. This allows subwavelength electromagnetic field manipulation but lacks a material base with high enough $|\chi^{(2)}|$. On the other hand, 2D-material photonics offers transition metal dichalcogenides (TMDs) compounds with remarkable ($|\chi^{(2)}|$) up to 10$^{-7}$pm/V for monolayer \ce{MoS2}, but struggles to maintain such high nonlinear response in most bulk TMD crystals or require additional photonic systems to enhance optical field in the vicinity of the monolayer flake. Luckily, this limitation can be overcome with a unique stacking in van der Waals (vdW) 3R-\ce{MoS2} crystals, which exhibit a higher refractive index ($n>4.5$) and $|\chi^{(2)}|$ in the near-IR compared to most of the all-dielectric materials. In this study, we present the first demonstration of a fully TMD bulk-based nonlinear resonant nanophotonics material platform that supports the transition from high-Q bound-states in the continuum (BIC) resonances in circular hole square lattice to quasi-BIC by symmetry breaking of the structure with wet etching, resulting in atomically sharp triangular hole square lattice. Such metasurfaces with significant second-order nonlinearity within the bulk of the structure, result in ultimate enhancement of efficiencies for all optically second-order nonlinear processes. We validate our concept through precise structure engineering, leading to a dramatic increase in the second-harmonic generation (SHG) from a resonant metasurface that supports qBIC. This concept paves the way for advancements in vdW resonant nonlinear nanophotonics toward optical information processing and the generation of quantum states of light.

Dielectric metasurfaces that combine high-index materials with optical nonlinearities are widely recognized for their potential in various quantum and classical nanophotonic applications. However, the fabrication of high-quality metasurfaces poses significant material-dependent challenges, as their designs are often susceptible to disorder, defects, and scattering losses, which are particularly prone to occur at the edges of nanostructured features. Additionally, the choice of the material platforms featuring second-order optical nonlinearities, $\chi^{(2)}$, is limited to broken-inversion symmetry crystals such as GaAs, GaP, LiNbO$_3$, and various bulk van der Waals materials, including GaSe and NbOCl$_2$. Here, we use a combination of top-down lithography and anisotropic wet etching of a specially stacked van der Waals crystal -- 3R-\ce{MoS2}, which exhibits both a high refractive index and exceptional $\chi^{(2)}$ nonlinearity, to produce metasurfaces consisting of perfect equilateral triangle nanoholes with atomically precise zigzag edges. Due to the geometry of the triangle, the etching process is accompanied by a transition from an in-plane $C_4$ symmetric structure to a broken-in-plane symmetry configuration, thereby allowing for the realization of the quasi-bound-state-in-the-continuum (q-BIC) concept. The resulting ultrathin metasurface ($\sim$ 20-25 nm) demonstrates a remarkable enhancement in second-harmonic generation (SHG) -- over three orders of magnitude at specific wavelengths and linear polarization directions compared to a host flake.

%, which are challenging to nanopattern with high precision

%The unique properties of 3R-\ce{MoS2} enable the combination of nonlinear and high-index metasurface concepts within an ultrathin ($\sim$ 20-25 nm) van der Waals material. 

% This process transforms initially fabricated rough circular holes in a square lattice into an array of triangular holes with atomically precise zigzag edges. 

%This work opens the door to nonlinear nanophotonics in high-quality van der Waals metasurfaces, at the interplay of atomically sharp edges, high refractive index, and extreme nonlinearities. 

%This innovative concept not only opens new avenues for nonlinear nanophotonics but also holds promise for applications in optical information processing and quantum light state generation. By pushing the boundaries of vdW resonant nonlinear nanophotonics, we are paving the way for transformative advancements in optical technologies and quantum photonics.

%\red{GZ: two main ideas - resonant SHG and BIC access(but separately). For fs laser BIC is too narrow}
}




\maketitle

Transition metal dichalcogenides (TMDs) form a class of 2D materials widely recognized for their remarkable optical and electronic characteristics~\cite{manzeli20172d}. Due to their unique crystalline structure and direct band gaps in the visible and near-infrared spectral range, 2H-TMD monolayers, with a general formula MX$_2$ (M = Mo, W and X = S, Se or Te), exhibit broken inversion symmetry, making them useful for applications in nonlinear nanophotonics and optoelectronics~\cite{mak2010atomically,splendiani2010emerging,wang2018colloquium,khan2022optical}. 
Large values of the second-order nonlinear susceptibility $|\chi^{(2)}|$ are essential for processes, like second-harmonic generation (SHG)~\cite{zhou2015strong}, sum- and difference-frequency generation (SFG, DFG)~\cite{yao2019continuous,wang2020difference}, optical parametric amplification (OPA)~\cite{trovatello2021optical}, spontaneous parametric down-conversion (SPDC)~\cite{dinparasti2018towards,guo2023ultrathin}, electro-optical effect (EO)~\cite{sun2016optical}, and others. Indeed, non-centrosymmetric TMD monolayers are characterised by impressive $|\chi^{(2)}|$ values, reaching, for example, for \ce{MoS2} $\sim 10^{-7}$ m/V~\cite{kumar2013second} (when expressed in bulk-equivalent units). Despite large $|\chi^{(2)}|$ values, TMD monolayers face a challenge in achieving optimal nonlinear conversion efficiencies for SHG and SPDC, $\eta$~\cite{dinparasti2018towards,marini2018constraints}, due to nanometric thicknesses and, therefore, wave propagation length ($\eta_{\mathrm{SHG}}=I_{2\omega}/I_{\omega} \propto |\chi^{(2)}|^2L^2$~\cite{boyd2023nonlinear}, where $L$ is the propagation length). This is in contrast to conventional phase-matched bulk crystals that are able to frequency-double light with efficiencies exceeding 50\%~\cite{ou199285}.

A promising strategy to increase the nonlinear conversion efficiency of TMDs involves integrating the monolayer onto a resonant metasurface spectrally aligned with the active material's peak $|\chi^{(2)}|$ response~\cite{bernhardt2020quasi,wang2019improved, kravtsov2020nonlinear,maggiolini2023strongly,sortino2024van}. Although this approach can enhance the nonlinear response, it does not directly address the volume constraints of TMD materials. Furthermore, this approach introduces additional complexities due to the involvement of semiconductor interfaces and strain effects during the exfoliation and integration of monolayers with resonant metasurfaces. These factors can result in spectral shifts of exciton resonances~\cite{sortino2019enhanced,benimetskiy2019measurement} and damping of nonlinear responses~\cite{kuhner2022radial,mennel2019second}, negatively impacting the coupled system's performance.

Hence, innovative approaches are required to unlock the full potential of various nanostructured materials for integrated nanophotonics and quantum technologies. In this regard, high-quality factor (high-$Q$) resonances in traditional dielectric nonlinear materials such as LiNbO$_3$~\cite{lu2019periodically,fedotova2020second}, BaTiO$_3$~\cite{karvounis2020barium}, GaAs~\cite{xu2019forward}, GaP~\cite{tilmann2023comparison} with considerably smaller $|\chi^{(2)}|$ values ($\sim 10^{-11} - 10^{-10}$ m/V) in comparison to monolayer TMDs are especially promising~\cite{smirnova2016multipolar, kivshar2018all, sain2019nonlinear}. Various designs including metasurfaces~\cite{lochner2018polarization,carletti2021steering}, waveguides~\cite{wang2017metasurface}, and whispering gallery modes (WGM)~\cite{kuo2014second} exhibit a dramatic efficiency increase of optical nonlinear processes including SHG and  SPDC~\cite{santiago2021photon}. The goal of this nonlinear all-dielectric nanophotonic concept is to identify materials with a high refractive index ($n$) that are robust to high-intensity excitation, have low optical losses ($k$), and high $|\chi^{(2)}|$ nonlinearity to efficiently confine electromagnetic fields within resonant nanostructures to generate a strong nonlinear response.

%Such a nonlinear all-dielectric nanophotonic concept aims to find materials with the highest refractive index ($n$), lowest optical losses ($k$), and highest $|\chi^{(2)}|$ nonlinearity to be able to effectively tightly localize electromagnetic field within the volume of resonant nanostructures of a smaller size and the material would not thermally suffer from highly intensive incident light.

\begin{figure}%[ht!]
    \centering
    \includegraphics[width=0.75\linewidth]{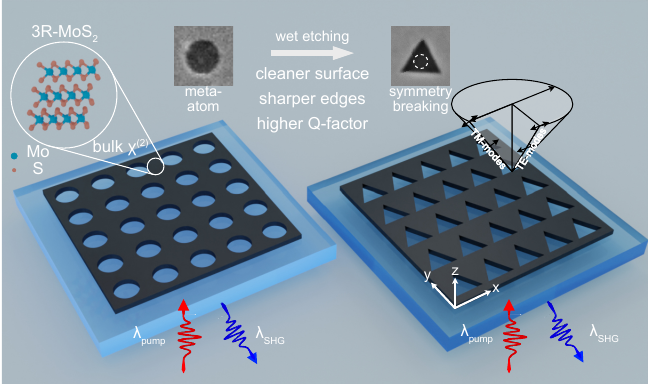}
    \caption{\textbf{Concept of broken-symmetry 3R-\ce{MoS2} metasurface.} Schematic of the wet etching process which forms atomically sharp equilateral triangular holes from the pre-etched circular holes of smaller size obtained \textit{via} reactive ion etching of 3R-\ce{MoS2} multilayer flake. Wet-etching provides atomically sharp edges, clean surface, and broken $C_2$ and $C_4$ symmetry, enabling the realization of q-BIC concept. Note that since wet etching consumes part of the material, it is necessary to start with a smaller initial circular hole diameter to achieve the desired size of the final triangular hole. Therefore, to achieve a comparable filling factor (ff) between circular and triangular metasurfaces, the initial circular hole size in the latter must be smaller in comparison to that of the former (see the dotted line representing the size of the original circular hole, as shown in the scanning electron microscopy image of the triangular nanohole).}
    \label{fig:TOC}
\end{figure}

A powerful contemporary concept to achieve optimal field confinement in nanophotonic structures is optical bound-state-in-the-continuum (BIC)~\cite{hsu2013observation, plotnik2011experimental}, which benefits from high-index dielectric materials~\cite{rybin2017high}. BICs feature electromagnetic field decoupling from the environment and, as a result, theoretically, an infinite quality factor, which, however, in real structures is limited by fabrication intolerances and material losses~\cite{hsu2016bound,koshelev2019meta}. Such high-$Q$ BIC resonances are challenging to excite from the far field due to optical reciprocity -- low optical leakage leads to weak optical excitation. By introducing a slight asymmetry in the system, one can efficiently excite these high-$Q$ resonances, at the expense of reducing their $Q$-factor~\cite{koshelev2018asymmetric}. Such a concept, often referred to in the literature as quasi-BIC (q-BIC), has great potential in a variety of nonlinear processes. It has been recently demonstrated that the SHG conversion efficiencies in structures featuring q-BICs can reach $\sim$ 0.1\%~\cite{carletti2018giant,liu2019high,koshelev2020subwavelength}, considerable SPDC~\cite{parry2021enhanced,santiago2022resonant,sharapova2023nonlinear}, and even high-harmonics generation~\cite{zograf2022high,zalogina2023high}.

Multilayer TMD materials are widely recognized for their high index ($n>4$) and extremely anisotropic ($\Delta n >1.5$) optical properties~\cite{green2020optical,munkhbat2022optical}, which makes them attractive for the realization of BIC concepts. Specifically, their use encompasses enhanced light-matter coupling~\cite{verre2019transition, munkhbat2023nanostructured}, leveraging q-BICs to reach the strong coupling regime~\cite{weber2023intrinsic}, Mie modes to enhance the nonlinear signal originating from the broken inversion symmetry at the surface~\cite{busschaert2020transition,panmairevealing}, and lasing in WGM disks~\cite{sung2022room}. However, conventional 2H phase bulk van der Waals crystals are centrosymmetric, resulting in a vanishing nonlinear signal~\cite{busschaert2020transition,panmairevealing}. In contrast, alternative TMDs, such as GaSe, NbOCl$_2$, 3R-\ce{MoS2}, and 3R-\ce{WS2}, which possess substantial $|\chi^{(2)}|$ nonlinearities even in bulk crystals ($\sim 10^{-10} - 10^{-9}$ m/V), have recently attracted research attention, particularly in terms of nonlinear optical and piezoelectric applications~\cite{zhou2015strong,gan2018microwatts,guo2023ultrathin,zhao2016atomically,shi20173r,xu2022towards,hallil2022strong,dong2023giant,zograf2024combining,weissflog2024tunable,qin2024interfacial,feng2024polarization}. However, to date, no high-index metasurfaces fabricated using these materials have been reported.  

% ,trovatello2023quasi - arxiv
%  \red{GZ: Davoyan's work~\cite{ling2024nonlinear}} 

In this work, we present the first demonstration of a bulk-$\chi^{(2)}$-active all-TMD metasurface supporting q-BIC resonances. Our high-index nonlinear 3R-\ce{MoS2} metasurface with a deeply subwavelength 20-25 nm thickness is able to significantly enhance the SHG signal -- more than three orders of magnitude compared to that of a host flake. We effectively control the optical losses of the BIC system through a wet etching process applied to convert a circular hole metasurface into the one composed of perfect equilateral triangles with atomically sharp edges~\cite{munkhbat2020transition}. This precise etching technique allows us to break the $C_2$ and $C_4$ in-plane symmetries, facilitating the excitation of the q-BIC resonances by far-field illumination. Furthermore, the ultrasharp nature of meta-atoms' edges enable the experimental realization of high-$Q$ optical resonances~\cite{kuhne2021fabrication}, as schematically illustrated in Fig.~\ref{fig:TOC}. Together with the recently developed wafer-scale fabrication method of the 3R phase van der Waals materials~\cite{qin2024interfacial}, this all-TMD nonlinear high-$Q$ concept holds promise to enhance the efficiency of a wide range of second-order nonlinear processes in deeply subwavelength nanophotonic settings.

%In this work, we reveal the first demonstration of a bulk-$\chi^{(2)}$-active all-TMD second-order nonlinear nanophotonic system supporting BIC and quasi-BIC states for enhanced SHG in a high-index nonlinear 3R-\ce{MoS2} metasurface of just $\sim 20$ nm thickness. We control the optical losses of the BIC system by gradual wet etching of the circular hole photonic slab array~\cite{hsu2013observation}, which supports the BIC state, into perfect equilateral triangles~\cite{munkhbat2020transition}. Such atomically sharp procedure allows to break in-plane C$_4$ and $C_2$ symmetries to effectively permit the excitation of the quasi-BIC state from the far-field illumination. Moreover, the wet-etching allows for a much cleaner sample surface and sharpness of the edges of the metaatoms, which is crucial for optical high-Q states~\cite{kuhne2021fabrication} (schematic of the idea is shown in Figure~\ref{fig:TOC}). As a result, we were able to achieve an improvement of more than 3 orders of magnitude improvement of the SHG intensity of the metasurface compared to an optimized thickness host flake. Such a compact TMD-based nonlinear high-Q concept should be attractive to any second-order optical nonlinear processes which are governed by $\chi^{(2)}$ tensor, providing the development of all-TMD nonlinear resonant nanophotonics.

%\section{Results and discussion}\label{sec:res}

\textbf{3R-\ce{MoS2} flake thickness optimization.}

We begin by optimizing the 3R-\ce{MoS2} material platform through analysis of the SHG signal as a function of both pump wavelength and the thickness of unpatterned flakes. This approach allows us to gain insight into how optical resonances within a metasurface can amplify the inherently efficient nonlinear response observed in bare flakes. Although similar studies have been previously reported, they focused on a specific or limited range of pump wavelengths~\cite{zhao2016atomically,xu2022towards,shi20173r}. Our work aims to expand this understanding by exploring a wider range of wavelengths and preparing a solid foundation for following metasurface experiments.

%We begin our study with the optimization of the 3R-\ce{MoS2} material platform by studying second-harmonic generation (SHG) intensity as a function of pump wavelength and height of the unpatterned flakes. With such a study, we would understand how effectively engineered optical resonances in the metasurface can enhance already very efficient nonlinear responses in bare flakes. Such studies were performed in several papers, however, they were limited to one or a few pump wavelengths~\cite{zhao2016atomically,xu2022towards,shi20173r}.

\begin{figure}%[ht!]
    \centering
    \includegraphics[width=0.99\linewidth]{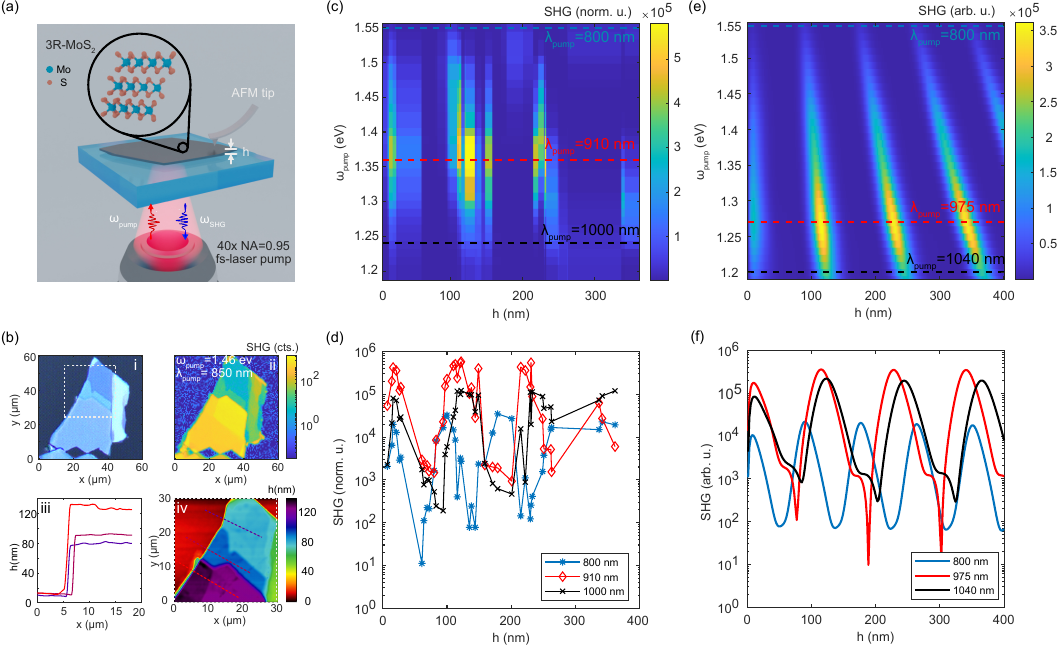}
    \caption{\textbf{Flake thickness optimization.} (a) Sketch of the SHG measurement of the exfoliated on glass substrate 3R-\ce{MoS2} flakes. The height of each flake was measured with AFM. (b) Standard study steps of the flake analysis: (i) Optical image of the flake of interest; (ii) SHG mapping of the intensity at a given pump wavelength; (iii) and (iv) AFM analysis of the flake height. The white dashed line in (i) corresponds to the AFM area of analysis in (iv). (c) Experimental SHG intensity of 3R-\ce{MoS2} flakes. Schematic of the experiment is shown in (a). Height $h$ of the flakes measured by AFM. Dashed lines correspond to cross-sections in (d). (d) SHG intensity cross-sections at given pump wavelengths. Markers represent experimental points. Solid lines connect neighboring points for visual perception. (e) Theoretically calculated SHG intensity of the 3R-\ce{MoS2} flakes using the TMM method. Dashed lines correspond to cross-sections shown in (f). (f) SHG intensity from flake thickness at given pump wavelengths. }
    \label{fig:material}
\end{figure}

Figure~\ref{fig:material} shows an experimental and theoretical study of the total intensity of the SHG as a function of a pump wavelength ($\lambda_\mathrm{pump}$ in 800 -- 1040 nm range or photon energy $\omega_\mathrm{pump}$ in 1.2 -- 1.55 eV range) and the height of the bare 3R-\ce{MoS2} flake exfoliated onto a glass substrate. The schematic of such a study is illustrated in Fig.~\ref{fig:material}a, where a tunable femtosecond laser is focused through an objective on the 3R-\ce{MoS2} flake surface from the glass substrate side. The SHG signal is then collected through the same lens and quantified by an avalanche photodiode (APD), provided that SHG spectra are free of other signals~\cite{zograf2024combining}. As shown in Fig.~\ref{fig:material}b, (i) an optical image of the flake is correlated with (ii) SHG mapping of the same flake at different pump wavelengths (example of a single wavelength SHG mapping at a single pump wavelength; SHG spectra, as well as SHG signal \textit{vs.} pump power are presented in Supplementary Information (SI), Fig. S1). Finally, (iii, iv) atomic force microscopy (AFM) mapping scans and profiles are measured to assess the relevant thicknesses. This approach allows us to collect SHG spectra for several flake thicknesses simultaneously.

%Each thickness of the flake was measured individually using AFM. 
%The dashed lines represent specific pump wavelengths of interest for which we provide an explicit comparison (Fig.~\ref{fig:material}d). 

The experimental result of the SHG intensity of the 3R-\ce{MoS2} flakes for different pump wavelengths, $\lambda_{\textrm{pump}}$, and flake thicknesses, $h$, is shown in Fig.~\ref{fig:material}c. Clear bands with high-intensity SHG are observed. They disperse toward increased thickness with decreasing pump energy, indicating resonant Fabry-Pérot (FP) behavior. The bright yellow parts of the map denote the maximum local SHG signal which is determined by both the resonant FP conditions and the spectral dependence of the experimental $\chi^{(2)}$ tensor.
Interestingly, the data show a narrow local SHG maximum at small $h$ ($\sim$ 20 nm), whose amplitude is smaller than that at larger thicknesses. 
The relative amplitudes and widths of the SHG signal are further compared in Fig.~\ref{fig:material}d for selected wavelengths (marked in Fig.~\ref{fig:material}c by dashed lines): $\lambda_{\textrm{pump}}$ 800 nm (blue) and 1000 nm (black) -- as the limits of the spectral range of the pump laser, and 910 nm (red) as the most resonant wavelength. This head-to-head comparison illustrates the sensitivity of the FP resonances to the exact value of $h$, providing up to three orders of magnitude enhancement in SHG intensity compared to off-resonant cases. 
%Moreover, at shorter pump wavelengths, we observe optical resonances at smaller values of $h$, which aligns with the FP nature of the enhancement. 

The theoretical analysis of the SHG intensity was carried out using the transfer-matrix method (TMM) of a glass/3R-\ce{MoS2} interface utilizing the $\chi^{(2)}$ tensor calculated in~\cite{zograf2024combining} and refractive index from ellipsometric measurements~\cite{munkhbat2022optical} (for more detailed information see Methods). Figure~\ref{fig:material}e presents calculated SHG for different heights of the flake (up to 400 nm) in the same energy range as in experiments. The dispersive FP resonances visible in the calculations match very well the experimentally observed dispersion. However, local SHG maxima within each FP band are slightly offset to the red versus the experiments, showing maxima at around 1.27 eV ($\sim975$ nm), where the theoretical $\chi^{(2)}$ tensor peaks~\cite{zograf2024combining}. The dashed lines for a cross-section comparison are selected at different pump photon energies due to a slight spectral mismatch between the experimental and theoretical $\chi^{(2)}$ tensors. However, the behavior seen in Fig.~\ref{fig:material}f is similar to that observed experimentally, with a $\sim$ 3-order of magnitude variability in SHG for a given pump wavelength.

A comparison of the experimental and calculated data underlines that FP interference is the main factor responsible for the observed dispersion of the SHG signal, while the $\chi^{(2)}$ tensor modifies the spectral intensity along each FP band. Furthermore, the calculated SHG signal shows saturation with increasing thickness of the 3R-\ce{MoS2} material for higher-order FP modes. This is most likely caused by the absorption at the second harmonic emission wavelengths, where $\lambda_\mathrm{SHG} = \lambda_\mathrm{pump}/2$. Indeed, the imaginary part of \ce{MoS2}'s dielectric function in the 400 -- 500~nm range exceeds 10~\cite{munkhbat2022optical}, preventing light emission from deeper layers of the TMD. While a significant amplitude of the second-order polarization exists throughout the bulk of the flake, only part of it near the surface can radiate efficiently into the far field, while the remainder is strongly attenuated. Hence, we conclude that $\sim$ 20 -- 25 nm thick 3R-\ce{MoS2} flakes are the most promising in terms of SHG emission per amount of material used in the studied spectral range.

\textbf{Metasurface with circular holes in a square lattice.}
An often-used design for a resonant metasurface that exhibits a high-$Q$ resonance is a square array of circular holes with a diameter $D$ and a center-to-center pitch $\Lambda$~\cite{hsu2013observation,liu2023boosting}. When certain symmetry conditions are fulfilled, such arrays may support so-called symmetry-protected and/or accidental BICs, depending on the angle of incidence and the exact arrangement of holes. The spectral range at which such a square array supports a BIC is determined by the material and geometrical parameters and it is critical to ensure they are chosen to facilitate both sharp resonances as well as efficient baseline SHG signal. The 3R-\ce{MoS2} flake thickness analysis indicates that there are several optimal thicknesses, which are candidates for metasurface nanofabrication. One may, in principle, aim at the second optimum, at $\sim$ 100 nm flake thickness with the strongest SHG signal or the thinner flake with approximately 50\% smaller SHG. In the former case, it may be possible that SHG emission will not be able to outcouple efficiently from a thick metasurface. Furthermore, thick samples may present difficulties in achieving precise nanofabrication. This would deteriorate the quality factor of the desired resonances, which are very sensitive to the roughness of the sample~\cite{kuhne2021fabrication}. On the other hand, localizing a BIC mode or other high-$Q$ resonances in structures with sub-20~nm thicknesses may be challenging due to an insufficient amount of the material. Balancing the above benefits and limitations, we aim to engineer metasurfaces supporting high-$Q$ optical resonances in multilayer 3R-\ce{MoS2} films of optimal thicknesses -- approximately $h\sim 20-25$ nm, Fig.~\ref{fig:material}~\cite{anthur2020continuous}. 

%One of the most frequently used and promising designs for a resonant metasurface that exhibits a high-Q state is a two-dimensional periodic quadratic lattice consisting of a circular hole with a diameter $D$ and a center-to-center pitch $\Lambda$ \cite{hsu2013observation, liu2023boosting}. This design can support bound states in the continuum (BIC). 
%As a BIC is an electromagnetic eigenstate with an ultimately high-quality factor; however, it is very challenging to excite using far-field excitation.

Symmetry-protected BICs have been realized in square lattices of circular holes in a SiN metasurface at normal incidence~\cite{lee2012observation}. These resonances feature ultra-high quality factors but are challenging to excite from the far field. To effectively excite a BIC, one typically introduces a symmetry break, which can be achieved by illuminating the metasurface at an oblique incidence~\cite{lee2012observation} or by rotating the meta-atoms within the metasurface~\cite{koshelev2018asymmetric}. To implement the former, we employ back-focal plane (BFP) imaging and spectroscopy using an oil immersion objective to access the reflection of the metasurface at various incidence angles and polarizations. Upon reaching the normal angle of incidence, the signal related to the BIC mode should vanish from the reflection spectrum, indicating the divergence of the quality factor, as expected for symmetry-protected BICs.

\begin{figure}%[ht!]
    \centering
    \includegraphics[width=0.99\linewidth]{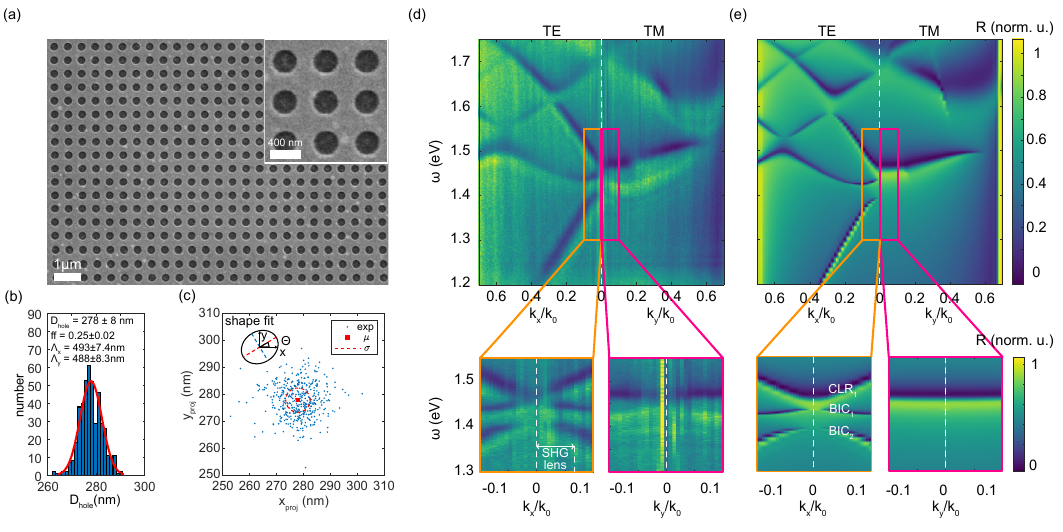}
    \caption{\textbf{Circular hole metasurface.} (a) SEM image of the circular hole 3R-\ce{MoS2} metasurface with $h = 27 \pm 1$ nm. The scale bar is 1~$\mu$m. The inset shows a zoomed-in image of the metasurface. The scale bar is 400 nm. (b) Histogram of hole diameters $D$ extracted from SEM images by analyzing the hole area assuming perfect circles. The mean value is $\langle D\rangle =278\pm 8$~nm; fill factor ff = 0.25$\pm$0.02; pitches $\Lambda_x$ = 493$\pm$7.4 nm, $\Lambda_y$ = 488$\pm$8.3 nm. The red line is a Gaussian curve fit. (c) Distribution semi-axes (minor, major) fitting of the holes assuming an elliptical cross-section. The projection of both semi-axes to the $x$- and $y$-axis provides the shape and size distribution of the circular holes in the metasurface. The red square marks the mean value of the circular fit (panel (b)), and the red dashed line denotes the mean value with the standard deviation. (d) Experimental BFP reflection spectroscopy of the metasurface for TE and TM modes with 2 close-ups. 'SHG lens' indicates the effective numerical aperture of the lens during SHG experiments in air. CLR -- circular lattice resonance, BIC -- bound-state-in-the-continuum. (e) Numerically calculated reflection spectra at different angles of reflection for TE and TM modes with similar close-ups. The parameters of the metasurface used in the calculation are: pitch $\Lambda = 500$~nm, $D = 262.5$~nm, $h = 23$~nm.}
    \label{fig:circ}
\end{figure}

The circular hole metasurface was fabricated using electron beam lithography (EBL) and reactive ion etching (RIE) on a multilayer 3R-\ce{MoS2} flake mechanically exfoliated from a macroscopic crystal (HQ Graphene, additional details are presented in Methods). Figure~\ref{fig:circ} shows an SEM image of a circular hole array and BFP reflection spectra from this array fabricated in a $27 \pm 1$ nm thick 3R-\ce{MoS2} flake. The thickness of the flake was measured using AFM.

The metasurface, presented in the SEM image Fig.~\ref{fig:circ}a, exhibits some shape imperfections and photoresist residues as a result of the non-ideal nanofabrication process and the usage of espacer (see Methods and Fig. S6). These features may negatively affect the overall performance of the metasurface~\cite{kuhne2021fabrication}. Deviations from circular shape are evaluated as shown in Fig.~\ref{fig:circ}c plot of the distribution of $x$- ($x_{\mathrm{proj}}$) and $y$-projections ($y_{\mathrm{proj}}$) of the major and minor ellipsoid axes. The mean value $\langle D\rangle$ (assuming perfect circular holes) is extracted from Fig.~\ref{fig:circ}b (see Methods and Fig. S3) and is plotted as a red square in Fig.~\ref{fig:circ}c, while the red dashed-line denoted the standard deviation of $\langle D\rangle$.

Figure~\ref{fig:circ}d presents the BFP reflection spectroscopy map of the circular holes metasurface (see Methods and Fig. S2 for additional details). The left and right sections correspond to transverse electric (TE) and transverse magnetic (TM) modes, respectively. We observe several narrow dispersive reflection minima corresponding to circular metasurface lattice resonances (CLR) and BIC modes. Specifically, due to the symmetry of the system, for circular nanohole metasurface, the TE and TM polarization modes are expected to align when reaching the normal angle of incidence. The region of interest in terms of angle of incidence is limited to $\sin{\theta} \approx 0.66$, which corresponds to the total internal reflection angle in the glass-air interface, calculated as $\theta = \arcsin{(n_{\mathrm{air}}/n_{\mathrm{glass}})} \approx 41.5 ^\circ$. The orange and red boxes indicate close-up images of the desired region measured in BFP reflection.

%The low-energy mode at $\sim$ 1.4 eV shows gradual narrowing and disappearance at normal incidence, as expected for symmetry-protected BICs. This is further confirmed by numerical simulations (see Fig.~\ref{fig:circ}e).

In the case of the TE mode (as shown in the orange box inset in Fig.~\ref{fig:circ}d), two symmetric reflection dips gradually thin down and eventually vanish near normal incidence at photon energies of approximately $\hbar \omega \sim 1.4$~eV and 1.45 eV, indicating the appearance of symmetry-protected BICs. In contrast, for the TM mode (indicated in the red box in Fig.~\ref{fig:circ}d), a single flat band exists within the entire range of angles in the inset, indicating that this band is not of a BIC character.

The BFP measurements are complemented by numerical calculations of reflection using the rigorous coupled-wave analysis (RCWA, see Methods for more details), see Fig.~\ref{fig:circ}e. This allows us to extract the optical parameters of the metasurface whose spectrum would match the experimental data. The parameters used in the calculations are $\Lambda = 500$~nm, $D = 262.5$~nm, $h = 23$~nm, and reasonably close to those extracted for the fabricated sample (see Methods). In addition, the calculations confirm our experimental observations, demonstrating a good agreement in terms of position and dispersion for both TM and TE modes, as is seen from the comparison of Figs.~\ref{fig:circ}d and e. Furthermore, our calculations confirm the CLR and BIC nature of the modes observed at normal incidence at $\sim 1.4 - 1.5$~eV. To conclude this section, while the reflection spectroscopy of the circular hole 3R-\ce{MoS2} metasurfaces shows promising performance, there remain several aspects to improve, particularly concerning the quality of individual meta-atoms and the precision of their edges.

\begin{figure}%[ht!]
    \centering
    \includegraphics[width=0.99\linewidth]{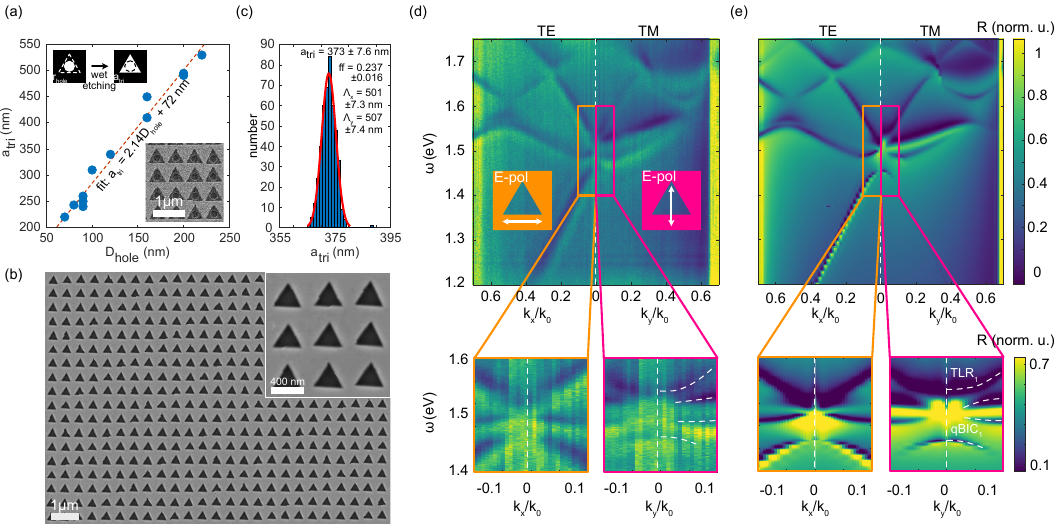}
    \caption{\textbf{Triangular hole metasurface.} (a) Experimental dependence of the equilateral triangular hole side size $a_\mathrm{tri}$ after wet-etching as a function of the diameter of the circular hole seed $D$. The linear fit function is $a_\mathrm{tri} = 2.14D + 72$ nm. The top inset is a schematic of a wet-etching process. The bottom inset is an SEM image of the triangular hole after wet-etching with a circular hole for reference. The scale bar is 1 $\mu$m. (b) High-resolution SEM image of the $h = 24.5 \pm 1.5$ nm thick metasurface. The scale bar is 1~$\mu$m. Inset is a close-up SEM image of the metasurface. The scale bar is 400 nm. (c) Triangular hole side size $a_\mathrm{tri}$ distribution analysis from the high-resolution SEM image in (b) using the threshold filter in ImageJ. The mean value is $a_\mathrm{tri}=373 \pm 7.6$~nm, ff = 0.24 $\pm$ 0.02; pitches $\Lambda_x$ = 501 $\pm$ 7.3 nm, $\Lambda_y$ = 507 $\pm$ 7.4 nm. (d) Experimental BFP reflection spectroscopy of the metasurface for TE and TM modes with 2 close-ups. TLR -- triangular lattice resonance, qBIC -- quasi-bound-state-in-the-continuum. (e) Numerically calculated reflection spectra at different angles of reflection for TE and TM modes with similar close-ups using following parameters: $a_\mathrm{tri}=375$ nm, pitch size $\Lambda=500$ nm, $h = 20$ nm.}
    \label{fig:tri}
\end{figure}

%\red{GZ: Betul, briefly the procedure and the main things to Methods, BK: done}.
\textbf{Quasi-BIC metasurface with triangular holes.} 

As a next step, we fabricate a 3R-\ce{MoS2} metasurface consisting of a square array of triangular holes using anisotropic wet etching of an initially pre-fabricated array of circular holes. This process follows the strategy reported previously for 2H-\ce{WS2} and 2H-\ce{MoS2} (for comparison between the etching of 2H- and 3R-\ce{MoS2}, see Fig. S4)~\cite{munkhbat2020transition,dewambrechies2023enhanced,polyakov2024top} and aims to accomplish two key objectives: (i) achieving ultrasharp edges of the triangular meta-atoms, and (ii) breaking the in-plane $C_2$ and $C_4$ rotational symmetry. The former aims at minimizing the scattering losses at the meta-atoms edges, while the latter is necessary to realize the q-BIC concept~\cite{koshelev2018asymmetric}. To ensure that the metasurface composed of triangular holes offers modes with similar dispersion in the same spectral range as those for the circular hole array, we keep the fill factor (ff) similar to that of the metasurface shown in Fig.~\ref{fig:circ}. Notably, since the material is consumed during wet etching, it is necessary to use a smaller diameter of the initial circular hole to achieve the required triangular hole size.

Figure~\ref{fig:tri}a demonstrates the typical dependence between the side length $a_\mathrm{tri}$ of the equilateral triangle hole and the diameter of the initial circular hole $D$. For a given thickness of the 3R-\ce{MoS2} flake ($h = 24.5 \pm 1.5$ nm), this dependence exhibits linear behavior, which here is fitted with the function $a_\mathrm{tri} = 2.14D + 72$~nm, where $D$ is the hole diameter in nanometers. The top inset in Fig.~\ref{fig:tri}a shows a schematic of the wet etching-induced shape transformation, while the bottom inset presents a typical SEM image of a triangular hole, initially fabricated as a circular hole through sequential EBL and RIE processes. 

It is important to note that material removal occurs in a specific manner: the edges of the triangular hole are oriented along the zigzag (ZZ) edge of the 3R-\ce{MoS2} crystal~\cite{dewambrechies2023enhanced}. Therefore, to achieve a triangular hole array in a perfect square lattice with one side of the triangle parallel to one of the axes of the unit cell, one must know the exact orientation of the ZZ and armchair (AC) directions of the crystal. Moreover, the strict correspondence between lattice orientation and crystalline coordinates plays a pivotal role in SHG experiments, as the $\chi^{(2)}$ tensor aligns with the crystalline axis, peaking in the AC direction, while optical resonances follow the metasurface lattice orientation. In this project, we aim to optimize both material properties and optical resonances; therefore, it is greatly beneficial that the orientations of nonlinearity and metasurface lattice are aligned.

Figure~\ref{fig:tri}b shows a high-resolution SEM image of a wet-etched metasurface with triangular holes. Notably, after the wet etching, the 3R-\ce{MoS2} surface appears to be significantly cleaner than in the case of circular holes, while the edges of the triangular holes are exceptionally sharp (approaching atomic sharpness, due to the self-limiting nature of the wet etching process~\cite{munkhbat2020transition,dewambrechies2023enhanced}). An analysis of the distribution of the triangular hole side lengths ($a_\mathrm{tri}$), which are extracted from the SEM image in Fig.~\ref{fig:tri}b, is presented in Fig.~\ref{fig:tri}c. Finally, we note that the anisotropic wet etching is able to reveal stacking faults that occasionally occur in 3R-\ce{MoS2} multilayers (see SI, Fig. S5); such faulty metasurfaces were excluded from reflection and SHG experiments. 

As with the circular hole array, we measure reflection spectra in BFP for the triangular hole metasurface. Unlike the former case, the unit cell with an equilateral triangular hole lacks in-plane $C_2$ and $C_4$ symmetries, therefore a metasurface based on such a cell is a promising candidate to allow the existence of the q-BIC state at normal incidence from the far-field excitation~\cite{koshelev2018asymmetric}. The collected experimental BFP spectra are plotted in Fig.~\ref{fig:tri}d for TE and TM polarizations with the same orientation as for the circular hole metasurface. Overall, the mode dispersion for the triangular metasurface shows similar behavior to the circular one, however, with a few notable exceptions. Most importantly, unlike the circular hole metasurface, in this case, a closer look reveals the presence of a reflection dip for the lower order mode even at normal incidence at 1.47~eV, indicating the presence of q-BIC. This normal incidence q-BIC is confirmed in the calculated reflection spectra for a metasurface with the following parameters: $a_\mathrm{tri}=375$~nm, lattice $\Lambda=500$~nm, $h = 20$~nm (see Fig.~\ref{fig:tri}e), which closely match the parameters extracted experimentally: $a_\mathrm{tri}=373\pm7.6$~nm, lattice pitches $\Lambda_x$ = 501$\pm$7.3 nm, $\Lambda_y$ = 507$\pm$7.4 nm, and flake thickness $h = 24.5 \pm 1.5$~nm (see Methods). Both experimental and theoretical results prove that the wet-etching fabrication is a powerful tool to break the in-plane symmetry of the meta-atom to access the q-BIC modes alongside reaching exceptional sharpness of the meta-atoms' edges.

% The error bars are estimated in a similar way to the case of the circular holes. 

\begin{figure}[ht!]
    \centering
    \includegraphics[width=0.8\linewidth]{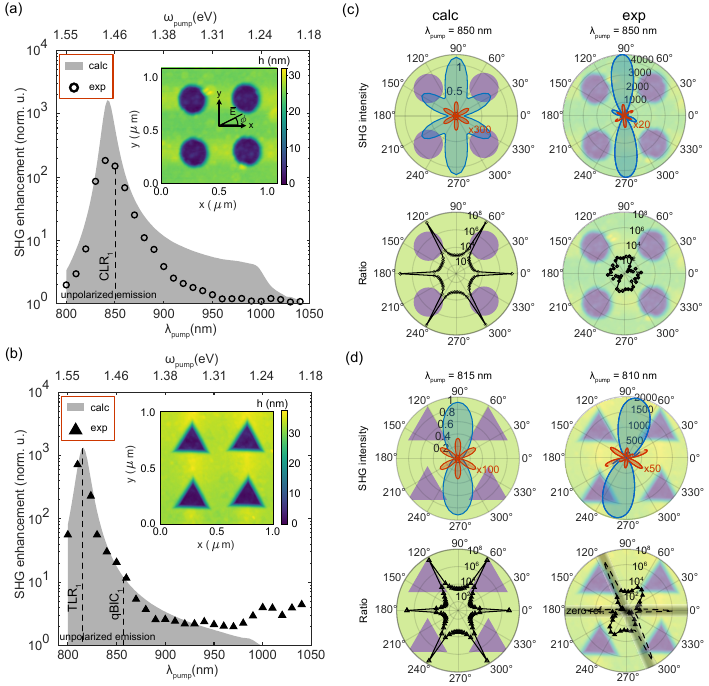}
    \caption{\textbf{SHG of circular and triangular hole metasurfaces.} (a) Circular hole metasurface SHG enhancement, SHG$_\mathrm{meta}$/SHG$_\mathrm{flake}$, \textit{vs.} pump wavelength $\lambda_\mathrm{pump}$. The pump polarization is linear along the $y$-axis. Detection has no analyzer. Circles correspond to experimental data, while the shaded area to numerical simulations. The inset depicts an AFM scan of the metasurface. (b) Triangular hole metasurface SHG enhancement,  SHG$_\mathrm{meta}$/SHG$_\mathrm{flake}$, \textit{vs.} the pump wavelength. The pump polarization is linear along $y$-axis. Detection has no analyzer. Triangles correspond to experimental data, while the shaded area to numerical calculations in Comsol. The inset depicts an AFM scan of the metasurface. 
    %(b) SHG enhancement (black dashed line) by a circular hole metasurface as a function of the linear polarization angle $\phi$ (left $y$-axis) at $\lambda_\mathrm{pump} = 850$~nm. SHG counts of the parallel to the linear polarization angle $\phi$ component for metasurface (blue) and the host flake (red) (right axis). 
    (c) Numerical (left column) and experimental (right column) polar plots of the SHG intensity (upper row) and SHG enhancement (lower row). SHG intensity is plotted in a linear scale. SHG enhancement (ratio) is plotted in a log-scale. The shaded background is the AFM image of the metasurface aligned with the polarization angles. 
    %(e) SHG enhancement (black dashed line) by a triangular metasurface as a function of the linear polarization angle $\phi$ (left $y$-axis) at $\lambda_\mathrm{pump}=810$~nm. SHG counts of the parallel to the linear polarization angle $\phi$ component for metasurface (blue) and the host flake (red) (right axis). 
    (d) Numerical (left column) and experimental (right column) polar plots of the SHG intensity (upper row) and SHG enhancement (lower row). SHG intensity is plotted in a linear scale. SHG enhancement (ratio) is plotted in a log-scale. The shaded background is the AFM image of the metasurface aligned with the polarization angles.}
    \label{fig:SHG_meta}
\end{figure}

\textbf{SHG enhancement in circular and triangular hole metasurfaces.}
After careful analysis of SHG in bare 3R-\ce{MoS2} flakes and BFP reflection of metasurfaces, which identified the existence of promising lattice resonance and optical BIC (q-BIC) modes in the 1.4 -- 1.6~eV range, we proceeded to study the SHG in those metasurfaces. Figure~\ref{fig:SHG_meta}a shows SHG enhancement spectra of the circular hole metasurface for a linearly polarized excitation along the $y$-axis (specified in the inset). Circular points are experimental SHG enhancement data, namely the ratio of the SHG intensity of the metasurface over the SHG intensity of the host flake at a given pump wavelength. We introduce this figure of merit -- SHG enhancement -- to quantify how the high-$Q$ modes can amplify the SHG signal of the metasurface compared to the SHG signal of the host flakes. Importantly, $\chi^{(2)}$ of 3R-\ce{MoS2} depends on the pump wavelength, therefore complicating direct comparison between metasurfaces with resonances occurring at different wavelengths. One can see that the circular hole metasurface features a clear resonant enhancement around 850~nm, which corresponds to a pump energy of 1.46~eV -- the same energy as the supported TM mode observed in the BFP spectra. The experimental SHG enhancement reaches $\sim$200-fold, while the numerically calculated one is almost an order of magnitude larger at the same pump wavelength (grey shaded area in Fig.~\ref{fig:SHG_meta}a). The numerically calculated SHG of the metasurface was carried out using the same parameters as the calculated reflection spectra in Fig.~\ref{fig:circ}c. The SHG of the host flake was calculated using the same thickness as the metasurface (for additional details about SHG calculations, see Methods). The mismatch in the magnitude of SHG enhancement between the calculation and experiment can be attributed to the surface and shape roughness of the circular holes, as one can see from the AFM image shown in the inset of Fig.~\ref{fig:SHG_meta}a, and from the SEM image in Fig.~\ref{fig:circ}a.

Similar experiments and numerical analysis were carried out for the triangular hole metasurface. Figure~\ref{fig:SHG_meta}b shows SHG enhancement of the triangular hole array with the incident polarization along the $y$-axis (along the height of the triangular hole and parallel to one of the axes of the unit cell). Similarly to the circular hole case, the spectral match is nearly perfect between the experimental SHG enhancement and the numerical one with the parameters extracted from Fig.~\ref{fig:tri}e, peaking at roughly $\lambda_\mathrm{pump}=810$~nm, which matches with the mode appearing in BFP around 1.53 eV in Fig.~\ref{fig:tri}d. However, in the case of the triangular hole metasurface, numerical and experimental data match not only spectrally, but also in the magnitude of the SHG enhancement -- $\sim$800-fold experimentally and $\sim$1400-fold for the numerical study. It is worth mentioning, that the numerically perfect circular and triangular hole arrays perform similarly in terms of the peak SHG enhancement ($\sim$1500-fold), however, experimentally we obtain substantially larger values for the triangular case -- $\sim$800-fold $\textit{vs.}$ $\sim$200-fold. We associate these results with the sharpness of the edges in triangular nanoholes and the cleanliness of the sample surface, as one can see in the AFM scan of the triangular hole sample with ultrasharp edges and a residual-free surface (Fig.~\ref{fig:SHG_meta}d inset).

After optimizing the pump wavelength for the circular hole metasurface at 850~nm, we perform polarization-resolved SHG measurements at this excitation wavelength (see Methods).
Figure~\ref{fig:SHG_meta}c shows experimental data of the SHG intensity of the circular hole metasurface (blue area) as a function of the angle $\phi$ characterizing the rotation of pump polarization with respect to the $x$-axis, and SHG of the nearby bare host flake (red line) at 850~nm. The black circles are the ratio between the SHG of the metasurface and the host flake indicating how efficiently the metasurface can enhance the SHG intensity at a given angle. One can see a clear reduction in the SHG enhancement around $\phi = 90^{\circ}$, which corresponds to the maximal SHG in both the host flake (along AC axis of the flake) and the lattice resonance of the metasurface.

The top left part in Fig.~\ref{fig:SHG_meta}c shows the calculated SHG intensity of the metasurface and the host flake at $\lambda_\mathrm{pump}=850$~nm overlaid with a schematic of the metasurface to illustrate the alignment of polarization-resolved SHG signal with respect to the metasurface axes. The data indicate perfect six-fold symmetry of the flake (red line) and 3 pairs of 2-fold symmetric features in the SHG of the metasurface with one dominant direction (blue line, dominant along $\phi = 90^{\circ}$ direction). Since the total SHG intensity is a product of the field enhancement and the nonlinearity of the material system, it is evident that the vertical direction is dominant for SHG in terms of both the AC direction of the flake and the lattice resonance of the metasurface. However, the appearance of non-zero SHG along the directions at which the host flake has theoretically vanishing SHG (ZZ-axis) provides substantial SHG enhancement (up to 8 orders of magnitude in calculations), which is plotted in the bottom left part of Fig.~\ref{fig:SHG_meta}c (black line).

The right part of Fig.~\ref{fig:SHG_meta}c shows experimental polarization-resolved SHG for circular hole metasurface in polar coordinates. The experimental data exhibits substantially less pronounced subpeaks (blue line, at $\sim$ $30^\circ$ and $330^\circ$), however, they appear along approximately the same direction as in the numerical calculations. Such discrepancy could be caused by (i) a slight misalignment of the crystalline axes and the metasurface unit cell and/or (ii) the metasurface not supporting high-$Q$ modes because of the shape and surface imperfections. The latter hypothesis is also confirmed by the modest SHG enhancement shown in the bottom right panel of Fig.~\ref{fig:SHG_meta}c (black line), where the maximum enhancement does not even reach 3 orders of magnitude along the ZZ direction where the reference flake exhibits minimum SHG.

%, suggesting an approximate alignment of the crystal axes with the metasurface axes, as in the theoretical scenario

The polarization-resolved analysis of the SHG enhancement in triangular metasurface is shown in Fig.~\ref{fig:SHG_meta}d. The top left panel shows a numerical calculation of the SHG intensity of the triangular hole metasurface (blue line) and the host flake (red line) as a function of $\phi$. In this case, the SHG intensity of the metasurface exhibits a single 2-fold symmetry feature along the triangular hole array lattice and the AC axis of the host flake -- $y$-axis. This is in agreement with the experimental result, shown in the right upper panel, however, the experimental polar plot exhibits a slight tilt in the peak position (by several degrees in a clockwise direction). This tilt is also observed in AFM (inset Fig.~\ref{fig:SHG_meta}b) and SEM (Fig.~\ref{fig:tri}b) images of the triangles and is likely caused by a slight misalignment of the crystalline axes of the 3R-\ce{MoS2} host flake with respect to the metasurface's unit cell, which in turn induces a slight symmetry break. Finally, the calculated SHG enhancement in the bottom panel of Fig.~\ref{fig:SHG_meta}d (left, black line) is in good agreement with the experimental data (right, black line) supporting the dominant direction of the enhancement along the ZZ direction of the flake. Notably, the experimentally measured SHG enhancement factor along $\phi \approx 60 ^{\circ}$ direction exceeds 10$^3$. The triangular nanohole metasurface presented here demonstrates performance that is considerably improved in comparison to its circular counterpart, making it a promising candidate for future nonlinear nanophotonics applications.

\textbf{Discussion and outlook}

In this work, we used the unique features of 3R-\ce{MoS2} to fabricate efficient ultrathin ($h$ = 20 -- 25 nm) nonlinear metasurfaces. First, we analyzed the thickness and wavelength dispersion of pristine 3R-\ce{MoS2} flakes for SHG in the $h$ = 0 -- 400 nm thicknesses and $\lambda_\mathrm{pump}$ = 800 -- 1040 nm pump wavelength (1.2 -- 1.55~eV) range. Second, we demonstrated the wet-etching fabrication of atomically sharp triangular hole metasurfaces with broken in-plane $C_2$ and $C_4$ symmetries, supporting q-BIC resonances. The atomic sharpness of the meta-atoms' edges leads to a substantial enhancement of the SHG signal, exceeding three orders of magnitude at specific resonant wavelengths and linear polarization conditions. Importantly, this enhancement is achieved with respect to the substantial SHG signal of the optimally thick host flake (corresponding to the first optimum thickness range, $h = 20 - 25$ nm, see Fig.~\ref{fig:material}). 

% Importantly, this enhancement is achieved relative to the already substantial SHG signal of the optimally thick host flake.s

We note that the metasurface resonances fall within the 800 -- 900 nm range, overlapping with the fundamental wavelength range of our tunable femtosecond laser. Consequently, in this study, we focus on optical resonances that overlap with the fundamental wavelength rather than the second harmonic. Nevertheless, our platform has the potential to support more complex configurations, where optical resonances could be, in principle, tuned to overlap with either the fundamental or second harmonic frequency, or even both simultaneously. 

Finally, our fabrication technique offers precise control over the photonic lattice, which is rigidly linked with the crystalline axes of 3R-\ce{MoS2}. The relative orientation between the crystal and metasurface axes introduces an additional degree of freedom for symmetry control, potentially enabling chiral harmonic generation, as predicted recently~\cite{koshelev2024scattering}. Furthermore, 3R-\ce{MoS2} metasurfaces may enable active tuning of the nonlinear emission response \textit{via} electro-optical modulation, carrier injection, and piezoelectric effects~\cite{hallil2022strong,dong2023giant}.  Thus, our work opens the door to nonlinear nanophotonics in high-quality van der Waals metasurfaces, at the interplay of atomically sharp edges, high refractive index, and extreme nonlinearities. 

%\textbf{Online content} Any methods, additional references, Nature Portfolio reporting summaries, source data, extended data, supplementary information, acknowledgments, peer review information; details of author contributions and competing interests; and statements of data and code availability are available at https://doi.org/xxx.

\textbf{Methods:}

\textbf{Smaple fabrication.}
 The nanofabrication steps were performed in the Myfab Nanofabrication Laboratory, MC2 Chalmers. A high-quality 3R-\ce{MoS2} crystal was purchased from HQ-Graphene and mechanically exfoliated into multilayer flakes. Multilayer flakes with the desired thickness were selected and transferred to 1-inch $\times$ 1-inch glass substrates (0.17~mm thickness) with the help of PDMS stamps (Gel-Pak, USA). PMMA was used as a positive resist for the electron beam lithography (EBL). A Raith EBPG 5200 (Germany) system operating at 100 kV accelerating voltage was used for the direct-writing EBL. The PMMA mask was nanopatterned with holes using 10~nA beam current. Oxford Plasmalab 100 system (RIE/ICP) was used with \ce{CHF3}/Ar to perform the dry etching step after EBL to transfer the pattern through the mask to the 3R-\ce{MoS2} flakes. The leftovers of the resist were removed by remover and, finally, the samples were washed with deionized water and gently blow-dried. For the circular hole arrays, the nanofabrication steps end here. However, for the triangular hole arrays, an additional step was performed to convert the initial circles to triangles. Conversion of the initial circles to atomically sharp triangles was performed using an anisotropic wet etching comprising an aqueous solution of hydrogen peroxide and ammonia, as described in more detail~\cite{battulga2020hexagon}. The nanofabricated devices were stored in a cleanroom environment till further experiments.
 
%\textbf{Sample characterization:}
\textbf{Atomic Force Microscopy.}
Atomic Force Microscopy (AFM) imaging was conducted using a Bruker ICON AFM system equipped with a Nanoscope 5 controller, operating in tapping mode. For the topography measurements in tapping mode, Bruker RTESP-300 AFM probes were employed.

\textbf{Scanning Electron Microscopy.}
SEM imaging of the circular and triangular hole metasurfaces was performed at Chalmers Materials Analysis Laboratory using an Ultra 55 microscope (Carl Zeiss). Espacer 300Z was spin-coated onto the samples as a conductive layer to reduce charging effects, as the samples were fabricated on a non-conductive glass substrate. An acceleration voltage of 1.7 kV was used for imaging both samples.

 The SEM images were analyzed using ImageJ software, where a monochromatic threshold filter was employed to distinguish the circular/triangular holes from the basal plane of the flake. The surface area $S$ of each circular/triangular hole was analyzed by ImageJ (450 holes in total). For the case of circular holes, the cross sections were fitted by an ellipsoid to obtain the major and minor axes, as well as the tilt angle between the major axis and the $x$-axis. The diameter distribution, which is plotted in Fig.~\ref{fig:circ}b, is derived from the surface area of each hole ($D$ = 2$\sqrt{S/\pi}$) assuming perfect circles. The analysis reveals pitch sizes of $\Lambda_x = 493\pm7.4$ nm, $\Lambda_y = 488\pm8.3$ nm, the mean diameter of the circular holes is $\langle D\rangle = 278\pm8$~nm, and the fill factor ff = 0.25$\pm$0.02, which is defined as $\mathrm{ff}=S/\Lambda^2$, \textit{i.e.} the ratio of the area of a circular hole to the area of the square unit cell.
 
 A similar analysis for triangular metasurface reveals the surface area $S_\triangle$ of each of the 450 holes, which, assuming equilateral triangles, results in the side lengths as $a_\mathrm{tri} = \sqrt{2S_\triangle/\sin{(\pi/3)}}$. The mean side length is $\langle a_\mathrm{tri}\rangle = 373 \pm 7.6$~nm. It is worth mentioning that the relative standard deviation of the triangular holes' size is smaller than that of circular holes, being 2\% and 3\%, respectively.

 We note that the error bars appearing in SEM image processing are dominated by systematic errors caused by the image resolution limitations of low-magnification images ($\sim$14 nm/pixel). Higher resolution SEM images, which contain enough nanoholes to perform meaningful statistics, were challenging to record in this case, due to significant charging effects, occurring because the samples resided on non-conductive SiO$_2$ substrates. Espacer was used to reduce the charging effects, however, it was not possible to fully avoid the issue. For the triangular nanoholes, several higher-resolution SEM images are shown in the SI (see Fig. S4), demonstrating that the edges of these holes are exceptionally sharp.

 %\red{The total error $\sigma_{\Sigma} = \sqrt{\sigma_{syst}^2 + \sigma_{stat}^2}$ of the hole size and pitch size estimation consists of two contributions: (i) systematic error $\sigma_{syst}$ - 7 nm (resolution of the SEM image - 14 nm/pixel); and (ii) statistical error $\sigma_{stat}$ - extracted from the histograms}.

%BK - details on SEM

\textbf{Optical reflection spectroscopy measurements.}
Linear optical reflection spectroscopy was carried out using a back-focal plane reflection optical setup. The setup consists of an inverted microscope (Nikon Eclipse TE2000-E) with an oil immersion objective NA = 1.49 (Nikon CFI Apo TIRF oil, MRD01691, 60$\times$). A laser-driven broadband light source (LDLS, EQ-99FC) was used to illuminate the samples from the glass substrate's side with an immersion oil (Nikon NF, n$_\textrm{oil}$ = 1.515). The immersion oil allows access to high NA directions of reflection with a theoretical limit of $\theta_\mathrm{max}= \arcsin{(\mathrm{NA}/n_\mathrm{oil})}$. The Fourier plane of the objective was imaged using a Bertrand lens and its images were recorded using a digital color camera (Nikon D300S). The spectra of the reflection collected in the back focal plane of the lens were recorded by a spectrometer (Andor Shamrock SR-500$i$, equipped with a CCD detector Andor Newton 920), while the fiber bundle consisting of 19 individual fibers allows us to collect 19 individual reflection spectra in the BFP simultaneously (the schematic of the setup is shown in the SI, Fig. S2).

%% The reflection spectra of the fabricated metasurfaces were studied using an inverted microscope by BFP imaging. Broadband illumination of the sample was achieved by a laser-driven white light source coupled to a high numerical aperture (NA) oil immersion objective, which focuses on the sample from the backside. A Bertrand lens is positioned in the back focal plane of the 60$\times$ NA = 1.49 objective, placed in front of the spectrometer. The reflected signal from the metasurface is collected using a fiber bundle containing 19 fibers that are coupled to the spectrometer. This fiber bundle simultaneously captures a range of angle-distributed reflection signals. The top channel is utilized to illuminate the sample with a low-intensity lamp for visualization with a camera.

\textbf{Second-harmonic generation spectroscopy measurements.}
The SHG of the metasurfaces was measured using an air NA~=~0.1 objective (Nikon CFI Plan Achro, 4$\times$) that focusing the tunable, 690 –- 1040 nm, Ti : sapphire femtosecond laser (MaiTai HP-Newport Spectra-Physics) with a $\sim$100 fs pulse duration and 80 MHz repetition rate. The low NA objective was chosen to mimic a near-normal incidence excitation, where angular dispersion of the metasurface resonances is expected to be weak (see BFP reflection data). The SHG signal was collected through a fiber and analyzed by a spectrometer (Andor Shamrock SR-500$i$, equipped with a CCD detector Andor Newton 920) for single spectra, and with an avalanche-photo-diode (APD, IDQ id100) for mapping using piezo-stage (Mad City Labs, MCL Nano-LP200) for precise positioning of the sample. The explicit SHG spectra, alongside SHG intensity \textit{vs.} input laser power demonstrating the slope = 2 in the log-log scale, thus proving the second-harmonic nature of the recorded signal, is presented in the SI (Fig. S1). SHG data presented in Fig.~\ref{fig:SHG_meta}a,b was obtained using a linearly-polarized laser beam and without an analyzer installed in the signal collection arm.

For the case of SHG experiments in Fig.~\ref{fig:material} we used an air NA = 0.95 objective with a correction ring (Nikon CFI Plan Apochromat Lambda D, 40$\times$) to compensate for distortion occurring by passing through the glass layer.

Polarization-resolved SHG data presented in Fig.~\ref{fig:SHG_meta}c,d was obtained using a linear polarization rotation with a $\lambda/2$ plate with $5^\circ$ steps. The SHG beam of the reflected beam passed the same sequence of $\lambda/2$ and linear polarizer as the incident light before being detected by the APD, therefore leaving only the component parallel to the incident linear polarization.

\textbf{Calculation of linear response of metasurfaces.} Reflection spectra of metasurfaces were calculated using the rigorous coupled wave analysis (RCWA) assuming a square periodic lattice. The geometrical parameters were taken from fabrication designs, structural characterization, and SEM images and then varied in a narrow parameter range to obtain good matching to experimental spectra. The permittivity of the semi-infinite glass substrate was assumed to be 2.25 and that of \ce{MoS2} was taken from ellipsometric measurements~\cite{munkhbat2022optical}. The resolution of the patterned \ce{MoS2} layer was \unit[2.5]{nm} and we took up to 250 terms for the plane wave expansion of the fields, which ensured converged results. For circular holes, we assumed azimuthal incidence along one of the principal axes of the square lattice. For the triangular pattern, we considered incidence both along the base as well as the height of the triangle.

\textbf{Calculation of 3R-\ce{MoS2} bare film SHG.} The SHG spectra of 3R-\ce{MoS2} flakes at normal illumination and collection were calculated using the transfer matrix method. The material parameters were assumed as above, while the nonlinear susceptibility was taken from electronic structure calculations reported in Zograf \textit{et al.}~\cite{zograf2024combining}. We calculated the electric field profiles in the \ce{MoS2} layer at the fundamental and second harmonic frequencies. The fundamental mode field profile was used to obtain the macroscopic nonlinear polarization. The nonlinear polarization was then multiplied by the mode profile at the second harmonic and an overlap integral was evaluated to yield the SHG signal at normal emission~\cite{obrien2015nonlinearFromLinear}.

\textbf{Calculation of metasurface SHG.}
Due to $D_{3h}$ symmetry of the 3R-\ce{MoS2}, the total SHG nonlinear polarization can be estimated through the following relation~\cite{boyd2023nonlinear,zu2022analytical,Psilodimitrakopoulos2024}:

\begin{equation} \label{eq:SHG}
    \textbf{P}(2\omega) = 
    \begin{bmatrix} P_x \\ P_y \\ P_z  \end{bmatrix} \propto 
    \begin{bmatrix} -2|\chi^{(2)}|E_x(\omega)E_y(\omega) \\ |\chi^{(2)}|(E_y(\omega)^2 - E_x(\omega)^2) \\ 0 \end{bmatrix},
\end{equation}

where $|\chi^{(2)}| = \chi^{(2)}_{yyy} = - \chi^{(2)}_{yxx} = - \chi^{(2)}_{xxy} = - \chi^{(2)}_{xyx}$ -- being the dominant tensor components due to symmetry constraints. The $\chi^{(2)}$ dispersion was taken from~\cite{zograf2024combining}. To reproduce results for polarization-resolved SHG polar plots for normal incidence of the pump along $z$-axis with just the parallel component of the SHG signal passing through the analyzer, we multiply the resulting SHG signal by the Jones matrix for a given linear polarization angle $\phi$:

\begin{equation} \label{eq:jones}
    \textbf{P}_{\phi}(2\omega) =  \begin{bmatrix} \textrm{cos}^2(\phi)& \textrm{cos}(\phi)\textrm{sin}(\phi) \\ \textrm{cos}(\phi)\textrm{sin}(\phi) & \textrm{sin}^2(\phi) \end{bmatrix} \begin{bmatrix} P_x \\ P_y \end{bmatrix},
\end{equation}

where $\textbf{P}_{\phi}(2\omega)$ indicates the SHG nonlinear polarization along angle $\phi$.

The SHG calculation of the periodic arrays (metasurfaces) was carried out in Comsol Multiphysics 6.1. The computational domain consisted of a rectangular box with a square cross-section along $xy$-plane. The size of the square section is defined by the pitch size of the metasurface $\Lambda$. The excitation port was placed on top of the computational domain and the excitation wave was set to propagate along the $z$-axis. The linear polarization of the excitation field was $E_x = E_0\cos{\phi}$ and $E_y = E_0\sin{\phi}$, where $E_0$ is the incident light intensity. The adjacent domain walls perpendicular to the $xy$ plane have a Flouqet periodicity for the $k$-vector. The computational steps consisted of two physical problems: (i) -- fundamental field and (ii) -- second harmonic. In step (i), the problem of the electromagnetic field distribution in the metasurface at 800 -- 1040 pump wavelengths is solved. The values of the $E_x (\omega)$ and $E_y(\omega)$ at the fundamental wavelength define the nonlinear polarization used in the second step (see Eq.~\ref{eq:SHG}). Then, in step (ii), the electromagnetic wave distribution problem is calculated at the second-harmonic wavelength. The total SHG intensity collected during the experiment is defined by the Poynting vector penetrating the upper surface of the computational domain. For polarization-resolved SHG calculations and SHG enhancement of the parallel component, we use the same computational model, however, the emitted SHG is then analyzed by taking into account the Jones matrix as described in Eq.~\ref{eq:jones}.

\backmatter

%\bmhead{Data availability} The principal data for this paper are available \textit{via} Zenodo at https://doi. org/yyy.

\bmhead{Acknowledgments}
G.Z. and BK are grateful to Dr. Adriana Canales for assistance with the optical back-focal plane setup. G.Z. thanks Ryo Mizuta Graphics for the optical components library for optical setups visualization. T.O.S. acknowledges funding from the Swedish Research Council (VR project, grant No. 2022-03347), Knut and Alice Wallenberg Foundation (KAW, grant No. 2019.0140), Chalmers Area of Advance Nano, 2D-TECH VINNOVA competence center (Ref. 2019-00068), and Olle Engkvist foundation (grant No. 211-0063). M.B. and T.J.A. acknowledge support from the Polish National Science Center \textit{via} the project 2019/34/E/ST3/00359.  W.W. acknowledges support from the Knut and Alice Wallenberg Foundation (KAW fellow, grant No. 2019.0231). Calculations were partially done at the ICM, UW (\#GC84-51). This work was performed in part at the Chalmers Material Analysis Laboratory, CMAL. This work was performed in part at Myfab Chalmers. 

\bmhead{Author contributions}
G.Z., B.K., W.W., and T.O.S. conceived the idea. G.Z. with the help of B.K. measured all linear and non-linear optical responses of the samples. B.K. with the help of A.P. fabricated all the samples. B.K. performed SEM imaging. A.A. provided AFM imaging of all samples. T.J.A. with the help of M.B. and G.Z. investigated numerically linear and nonlinear optical responses of the samples. T.J.A. and T.O.S supervised the theoretical and experimental parts of the project, respectively. G.Z., B.K., and T.O.S wrote the manuscript with the contributions of all co-authors.

%\section*{Declarations}


%\bibliography{Main_bib}% common bib file


\end{document}